# EQL -- an extremely easy to learn knowledge graph query language, achieving high-speed and precise search

Han Liu    Shantao Liu
Beijing Lemon Flower Technology Co., Ltd.

2020.3.15

Abstract:

EQL , also named as Extremely Simple Query Language , can be widely used in the fields of knowledge graph, precise search, strong artificial intelligence, database, smart speaker, patent search and other fields. EQL adopts the principle of minimalism in design and pursues simplicity and easy to learn so that everyone can master it quickly.

In the underlying implementation, through the use of technologies such as NLP and knowledge graph in artificial intelligence , as well as technologies in the field of big data and search engine, EQL ensure high-speed and precise information search, thereby helping users better tap into the wealth contained in data assets.

EQL language and λ calculus are interconvertible, that reveals the mathematical nature of EQL language,and lays a solid foundation for rigor and logical integrity of EQL language.

In essence, any declarative statement is equivalent to a spo fact statement. The mathematical abstract form of the spo fact statement is s: p: o (q1: v1, q2: v2 ..... qn: vn), s:p:o is  subject:predicate:object, or entity : attribute : attribute value. (q1: v1, q2: v2 ..... qn: vn) are called modifiers to express more details, n is 0 or a natural number; any question statement is equivalent to one or more spo question statements, the basic form of each spo question statement is to replace any symbol of the spo fact statements with the ?x, for example, s: p:?x (q1: v1, q2: v2), or s: p: o (q1:?x, q2: v2, q3: v3).

The above proposition, which can be referred to as Alan Liu Theorem, is the theoretical basis of the EQL language. We would like to pay tribute to the great Godel Incompleteness Theorem and Church-Turing Theorem with Alan Liu Theorem.

Hundreds of billions of spo fact statements can constitute a knowledge graph with the world's commonsense. The essence of EQL is to expand the spo question statement so that people can easily access the information in the knowledge graph.

The EQL language and a comprehensive knowledge graph system with the world's commonsense can together form the foundation of strong AI in the future, and make up for the current lack of understanding of world's common sense by present AI system.

EQL language can be used not only by humans, but also as a basic language for data query and data exchange between robots.

1. Introduction

Since human society entered the information age, the data that humans possess has grown exponentially. The development of the Internet and mobile Internet in particular has accelerated the trend of data explosion. However, the language for people to query data has not improved for many years. The most commonly used data query language is still the SQL language proposed by Boyce and Chamberlin in 1974. SQL language is a professional relational database query language, but except programmers and database administrators, only a few users can master SQL.

In recent years, let by Google, great progress has been made in the field of knowledge graph, and the knowledge and data in the knowledge graph have greatly improved in quantity and quality. However, the query language for knowledge graph is still an important bottleneck to user. The SPARQL language that commonly used in the field of knowledge graph was first introduced in 2008 , and its complexity is basically equivalent to SQL language. The Cypher query language for graph databases , which came out in 2011 , is almost as difficult as SPARQL. It is still difficult for ordinary people to master.

The EQL(Extremely Simple Query Language) discussed in this article , can be widely used in the fields of knowledge graph, precise search, strong AI, database, smart speaker, patent search and so on. In the design of EQL, the principle of minimalism is adopted, and we pursue simplicity and easy to learn ,so that everyone can master it quickly.

On the bottom layer, through the use of technologies such as NLP natural language understanding and knowledge graph in artificial intelligence , as well as technologies in the field of big data and search engine, EQL ensure high-speed and presice information search, thereby helping users better tap into the wealth contained in data assets.

For the convenience of non-professional reader, this article try to put the simple content in the previous chapter and the difficult content in the back of the article.

To pay tribute to the tradition that SQL pronounced as sequel, EQL pronounced as equal , in international phonetic alphabet as IPA [ i : kw ə L ] . EQL symbolizes human's pursuit of freedom and equality, and helps ordinary people to get equal data empowerment in the information age.

2. The basic syntax of EQL

Let's start to learn EQL with some simple examples . In each case, we will first give a query question described in natural language, then an EQL statement followed by " EQL :" ,then the returned result followed by "ANS : ".

Case 1: understanding spo statements
  Question : Who won the 2016 Nobel Prize in Literature?
  EQL: ?: award : Nobel Prize in Literature (Date : 2016 )
  ANS: Bob Dylan

In the knowledge graph system that support EQL, a piece of fact data is stored in the following abstract form:
   s: p: o (q1: v1, q2: v2, q3: v3, ... , qn: vn)
  Such an abstract form is called a spo statement .

All the spo statements are combined into the knowledge graph system that we query. The comprehensive knowledge graph system may contain tens of billions of spo statements, and store the commonsense of most subjects created by human being. EQL language and comprehensive world's commonsense knowledge graph system can together form the foundation of future strong AI, and make up for the current lack of understanding of world's common sense by present AI system.
EQL language can be used not only by humans, but also as a basic language for data query and data exchange between robots.

The underlying data storage system supporting EQL language may have multiple options, such as SQL databases, RDF databases, graph databases, MongoDB and other NoSQL databases, and even character files or Excel files. After proper processing and conversion, and development of interface software, the existing data (such as patent data, geographic information data, e-commerce data, enterprise ERP system data) can also be transformed into a data source for EQL language.

Example: The fact is "Bob Dylan won the 2016 Nobel Prize in Literature with a prize of SEK 8,000,000 "

This fact is stored in the knowledge graph system as spo statements as follows:

Bob Dylan : award : Nobel Prize in Literature
 (date : 2016 ,  prize : 8000000 SEK )

among them:
    s = Bob Dylan , p = award , o = Nobel Prize in Literature
    q1 = date, v1 = 2016
    q2 = prize, v2 = 8000000 SEK

SEK stand for Swedish krona.

In the spo statement, s: p: o are the subject, predicate, and object, such as, "Abraham Lincoln: position held : US President".
Or  s:p:o are entity, property and property value, such as " Earth : radius : 6371 km " .

In the knowledge graph, all nouns can be used for subject as s, be collectively called entity. An entity can be everything, it can be a specific thing, such as a specific person, animal, plant, celestial body, mountain, etc., or it can be an abstract thing, such as time, space, concept, event, attribute, official position, formula, theorem, etc.

All facts of the same entity (such as "George Bernard Shaw" ) are combined to form the knowledge card of this entity. User can click the link of the knowledge card number to bring up the knowledge card for viewing. You can download the knowledge card in PDF,
Word or Excel format.  For a set of knowledge cards, you can select several properties of these entities,and generate Excel files for download.

The entity that can act as p is a verb or a property of entity, be collectively called property . Such as "position held" in
" Abraham Lincoln : position held : US President "  , or
" radius" in "Earth : radius : 6371 km" .

Regarding the encoding method of the entity, the encoding of the property is recommended to start with the letter p , where p stands for property . The encoding of all other entities, it is recommended to start with the letter e , where e stands for entity .

Object (or property value) can be an entity, such as "US President" .Or object can be specific data, such as "16410 sq km" in
" Beijing: area :16410 sq km ".

A spo statement represents a fact, and s: p: o is the main description part of this fact, which is called an spo clause. The parts (q1: v1, q2: v2, q3: v3, ...... qn: vn) are called modifiers and are referred to
as qv clauses. Modifiers illustrate the details of this fact. q1: v1 or qn: vn is called a modifier, and each fact can have 0- n modifiers (n is a natural number).

When n is 0 , the spo statement appears as a simple pattern of s: p: o . The corresponding EQL statement can be:
  ?: p: o
     EQL:  ? : Capital : Beijing
     ANS:  China
  s:?: o
     EQL: Bob Dylan :?: Male
     ANS: gender
  s: p:?
     EQL: bird : good at :?
     ANS: flying

The above EQL statements are used to query subject, predicate, object, or entity, property, and property value. Above EQL statements are so simple, I believe even children in kindergarten can master it.

These EQL statements can be further simplify to:
  ?po
     EQL:? Capital Beijing
     ANS: China
  s?o
     EQL: Bob Dylan ? Male
     ANS: gender
  sp?
     EQL:Bird good at ?
     ANS:flying

Now the reader can realize why  EQL is called as Extremely Simple Query Language.

Based on natural language understanding in AI, EQL system can do automatic segmentation, automatically segment "Bird good at ? " into "Bird : good at :? ",  like the s: p: o standard form.

To get the basic form of EQL statement, you can change any part of spo statement s:p:o (q1: v1, q2: v2, q3: v3, ..... qn: vn ), replace it with '?', EQL system will search the knowledge graph accurately,and return the calculation result of '?'.

For example : (To see some data supporting the query in this article, please refer to Appendix 2 )

Case 2  ask entity s in spo statement
   Question : Who won the Nobel Prize in Literature in 1925 ?
   EQL: ?: award : Nobel Prize in Literature (Date : 1925 )
   ANS: George Bernard Shaw ehm001001

Note: ehm001001 is the entity number of "George Bernard Shaw" in the knowledge graph system. Clicking on this number,system will display the knowledge card about "George Bernard Shaw" , providing rich information similar to Wikipedia.

Case 3  ask property p in spo statement
   Question: What is the relationship between George Bernard Shaw and Dublin?
   EQL: George Bernard Shaw : ? : Dublin
   ANS: place of birth     p01000100

Note: Such a query can help users easily find the name of property  with an example

Case 4  ask object o in spo statement
   Question: Where was George Bernard Shaw born?
   EQL: George Bernard Shaw : city of birth : ?
   ANS: Dublin    ep1900101

Note: Click ep1900101, you will find this Dublin, the Irish capital Dublin, is not the United States town Dublin ( Dublin CA)  in the east of San Francisco Bay Area . Place and person often have the same name. Entity code is given in returned results to ensure the accuracy

of EQL answer. This is one of the reasons why EQL search is can do precise search.

Surely, such precise search should be based on the accurate semantic understanding and intelligent concept comparison of data when the knowledge graph is established. During data cleaning and data importing, some advanced technology in AI should be used,such as NLP natural language understanding and BERT in deep learning .

   Attentive readers may have noticed, the p in EQL statement is  " city of birth " , rather than " place of birth " , EQL can also perform the correct result, because in the knowledge graph, some aliases have been stored for the property " place of birth " .
   The corresponding spo statement is as follows:
      place of birth : alias : city of birth
      place of birth : alias : City of  Birth
      place of birth : alias : BirthCity
      place of birth : alias : BirthPlace
Using any alias to construct the EQL statement, you can get the correct result.

Case 5  ask some details of  of qv clasue in spo statement
  Question: What is the prize for George Bernard Shaw's Nobel Prize in Literature?
  EQL : George Bernard Shaw : award : Nobel Prize in Literature (prize : ?)
  ANS: 118165 SEK

Note: If this EQL statement is written in the SQL database query language, there are many ways to write it according to the design of the database structure. The following is a possible version.
       SQL: select prizeAmount from person
              where personName = 'George Bernard Shaw'
              and award = 'Nobel Prize in Literature'
    Under normal circumstances, to write the correct SQL statement, users need to understand the structure of many tables in the database and the names of each field. If you want to implement the encyclopedia knowledge database, in addition to the person table in the database , there will be many tables such as country , organization , place , event , product , animal, etc., which will soon exceed the memory capacity of ordinary people.

Case 6 ask all details of qv cluase in spo statement
  Question: What is the details that George Bernard Shaw won the Academy Award for Best Screenplay?
  EQL:George Bernard Shaw : Award : Academy Award for Best Screenplay (?)
  ANS:  (  Winning work: Flower Girl,
         Date: 1939 ,
         Related items: The 11th Academy Awards)

Note: In this case,  all qv clause will be the result, which contains all the details.

Case 7 get multiple rows of query results
  Question: Who has won the Nobel Prize in Literature ?
  EQL : ? : Award : Nobel Prize in Literature
  ANS: Geroge Bernard Shaw    ehm001001 ,
       Bob Dylan              ehm001002,
       Kazuo Ishiguro         ehm001003
       ...

Note: When there is a lot content in the returned result, EQL first returns 50 rows of data. The user can continue to click to get all the data or the next 50 rows of data. First returns 50 rows of data , is to allow users to get fast response.

3. Advanced Syntax of EQL

Case 8  get multiple rows of results and sort them
  Question: Who has won the Nobel Prize in Literature ? Please sort in order of the time
  EQL: ?x: award : Nobel Prize in Literature ( Date :?y)
          \order by  ?y desc
  ANS: x = Peter  Handke    ehm001009       y = 2019,
       x = Olga Tokarczuk  ehf001008       y = 2018,
       x = Kazuo Ishiguro   ehm001003       y = 2017,
       x = Bob Dylan        ehm001002       y = 2016,
       .......

Note: You can use ? x or ? x ,? y ,? z represents the variable to be queried. When there is more variables, you can use x, y, z1, z2, z3, .... to represent, in principle, the total number of variables does not exceed 30.

In above EQL statement, "\order by ?y desc" called \order sorting clause, 'desc' represents descending order.
EQL language can automatically identify the numer and unit in the content of knowledge graph.For example, the date content " 5671  km" can be automatically analyzed to be digital "5671 " and length unit "km" .
When sorting the content of time property such as "date" , descending means that the event occurs later is ranked first.
'asc' stands for ascending order. If no order is specified, they are sorted in ascending order.
When writing the sort clause of EQL , it is recommended to start a new line and indent it, to facilitate easy comprehension.

Case 9 get multiple rows of query results and group them

Question: Who has won the Nobel Prize in Literature ? Please group according to the nationality of the winners.
EQL:? x: award: Nobel Prize in Literature ( Date :? y)
       \group by  ? x.nationality
       \order by  ?y asc
ANS: group x.nationality = United States

       ...
       x = Eugene  O'Neill     ehm001021   y = 1936 ,
       x = Pearl S. Buck       ehf001122   y = 1938 ,
       x = William Faulkner    ehm001023   y = 1949 ,
       x = Ernest Hemingway    ehm001024   y = 1954 ,
       ...
    group x.Nationality = France
       x = Romain Rolland      ehm001025   y = 1915 ,
       x = Anatole France      ehm001026   y = 1921 ,
       x = Andre Gide          ehm001027   y = 1947 ,
       ...

Note: "x.nationality" is an example of the s.p expression in
spo statement.  s.p is called a property expression and represents the property p of the entity s . For example, if "x = Ernest Hemingway" , then "x.nationality = United States".

When the result of s.p is an entity, you can continue to connect new property in series. The property expression formed in the way is s.p1.p2.p3 .... pn, and the maximum pn does not exceed p8.

For example, if "x = Ernest Hemingway ",
    then  "x.nationality.capital = Washington".

When returning rows of data exceeds 100000, sorting and grouping so many data may be more time-consuming, EQL system may tell user  the progress of calculation and estimated time remaining.

Case 10 get multiple rows of result and filter results with Filter clause
    Question: In United States,Who has won the Nobel Prize in Literature since 1940 ?
    EQL:   ? x: award: Nobel Prize in Literature (date:?y)
            \filter  ?x.nationality = United States
            \filter  ?y> 1940
    ANS:   x = William Faulkner     ehm001023   y = 1949 ,
           x = Ernest Hemingway    ehm001024   y = 1954 ,
           ...

Note: The clause begin with \filter is called filter clause. There can be multiple filter clauses. There is a logical AND relationship between them. What this meant is, the result obtained by the first filter clause,should be given to second filter clause to be filtered continuously, and so on.

The comparison operators in EQL include:
    =    Equal
    ! =  Not equal
    >    Greater than
    > = Greater than or equal
    <    Less than
    <=  Less than or equal

Case 11 get multiple rows of  results and filter them with string pattern matching

Question: Are there someone named " William" among the winners of Nobel Prize in Literature? What year did he win ?
EQL: ? x: award : Nobel Prize in Literature (date :?y)
   \filter ? x \match   '%William% '
ANS:  x = William Faulkner    ehm001023   y = 1949 ,
      x = William Golding     ehm001067   y = 1983 ,
      .......

Note: \match means  to match a string,it followed by a string matching template, such as '%William%' . '%' can match 0-n characters, that is, any string or empty string. If you want to match "William" at the beginning or the end of string, the matching template can be written as '%William' or 'William%' .

Another string match symbol is '_',  '_' means to match a character, which could be an English character, a Chinese character, or any unicode character.

For example, the matching template 'William G_lding' can match 'William Golding'. Matching template '赵_龙' can match '赵子龙', '赵小龙', etc.

In EQL string matching template, only two matching symbols '%' and '_' are used, that once again reflects the minimalist design principle of EQL.

In chinese Taoism book 《Tao Te Ching》, Lao Tzu wrote:
   "Out of  Spirit,One is born;
    Out of  One,Two;
    Out of  Two,Three;
    Out of  Three,the created universe."

In the small world of string matching, '%' and '_' are the Spirit and One. The essence of the Universe is also Spirit and One. May the Spirit and you are One, May the Force be with you.

If you want to match a string that contains '%'  or '_' in itself, you can use '\%' for '%' , use '\_' for '_'.

For example:
    EQL : ?x:instance of :song
          \fileter ?x  \match   '%99\%qualified%'

    ANS:  x= 99% qualified boyfriend   e03301003

Similarly, if there is ':' or  '?' in the content of spo statement, you can use '\:' for ':', use '\?' for '?' . For example, there are two books,  'What is life? The Physical Aspect of the Living Cell' and 'Central Asia : Culture on Horseback' . When querying the author, you can write EQL statements like this:

   EQL :What is life\? The Physical Aspect of the Living Cell: author:?
   ANS: Erwin Schrödinger  ehm001097

   EQL : Central Asia \: Culture on Horseback:author:?
   ANS: Yingjie Xiang   ehm021038

To facilitate user input, the '?' or ':' in the EQL statement can be entered under the English input method, it can also be entered under the Chinese input method.

Case 12 use \and,\or, \not to implement logical operation
   Question: Who won both the Nobel and Academy Awards?
   EQL: (?x: award) : instance of: Nobel Prize
        \and
     (?x: award) : instance of: Academy Award
  ANS: George Bernard Shaw  ehm001001,
      Bob Dylan            ehm001002

Note:
   a.EQL can use \and , \or , \not to perform logical AND, OR, NOT calculations, and use \true and \false to represent logical truth and false.
   b. the expression (s:p),like (?x: award), represented to calculated (s:p) use s and p in a spo statement, (s:p)=o where s,p,o satisfied (s: p: o) in the knowledge graph, and (s:p) can be the subject of subsequent spo statements in the EQL query.

   c. Here we briefly describe the logical calculation process for a row of spo statement data
     Suppose, when x = George Bernard Shaw, one line of spo fact data is:
       George Bernard Shaw : award : Nobel Prize in Literature
         (Date: 1925 ,
          prize: 118165 SEK)

The calculation process based on this row of data is as follows:
   (?x :award): instance of: Nobel Prize
   = (George Bernard Shaw: award): instance of: Nobel Prize

   = Nobel Prize in Literature: instance of : Nobel Prize

   = \true
The result of final step of the calculation is logical \true,because in the knowledge graph,there is following spo statement data:
   Nobel Prize in Literature : instance of: Nobel Prize

The logical " \true " calculated in this line continues to participate in subsequent logical calculations. Finally, by querying the relevant data in the knowledge graph, the " x = George Bernard Shaw" and ""x = Bob Dylan " is a result that meets the query conditions.

d. In Wikidata knowledge graph, the SPAQL statement used to do the same query is as follows:
 #People that received both Academy Award and Nobel Prize
   SELECT DISTINCT ?Person ?PersonLabel WHERE {
     ? person wdt: P166 / wdt: P31? wd: Q7191.
     ? person wdt: P166 / wdt: P31? wd: Q19020.
     SERVICE wikibase: label {
       bd: serviceParam wikibase: language "[AUTO_LANGUAGE], en".
     }
   }
To understand the above SPARQL statement, you need to understand that in Wikidata , P166 represents  property 'award', P31 represents property 'instance of ' ,  Q7191 represents the Nobel Prize, and Q19020 represents the Academy Award. It can be seen that the EQL language is significantly more concise and understandable than SPARQL .

SPARQL queries in Wikidata can achieve accurate search, but at the expense of losing convenience for users to learn and use. EQL language is easy to learn, and also maintain precise of search.

Case 13   reason for using \and to represent logical AND
   Question: Who is the author of "Pride and Prejudice" and "Sense and Sensibility"?

EQL: Pride and Prejudice:author :?x
   \and
   Sense and Sensibility: author :?x
ANS: x = Jane Austen    ehf001035

Note: In the above EQL statement, there are three 'and' , '\and' represents a logical operation at a glance.

The reason EQL uses '\and' instead of 'and' or '&' for logical AND calculations is because in spo statement, a lot of content will have 'and' or '&' . Using '\and' to represents logical AND , can greatly reduce the logic ambiguity.

EQL uses \or, \not, \true, \false for logical and semantic symbols, as reason similar to the case of \and .

Case 14   handling of duplicate names of entities such as person or place names
 Question: Which writers were born in Dublin?
 EQL : ?x: occupation :writer
         \and
      ? x: place of birth : Dublin
 ANS: George Bernard Shaw ehm001001,
      James Joyce           ehm001088,
      ...

   EQL duplicate name report:
     1. Dublin_Ireland   default
     2. Dublin_California
   The above query results is calculated by the default value
       "Dublin = Dublin_Ireland"

Note: The above "EQL duplicate name report:
     1. Dublin _ Ireland default
     2. Dublin_California
     ......"
  It is also part of the EQL query results.

People or places often have the same name. If one of them is obviously better known, we can take it as the default option. For example, usually we use Dublin to represent Dublin of Ireland, we are can use

"Dublin_California" to represent Dublin in California. When place names are duplicated, and we usually distinguish them by adding a higher-level administrative area to which they belong.

We usually distinguish people by occupation, for example:
    1. Yao Ming, Yao Ming_basketball player   default
    2. Yao Ming, Yao Ming_composer
  If we query the situation of basketball player Yao Ming, we use " Yao Ming" directly , such as:
    EQL: (Yao Ming : wife): graduation school:?

  If we query the situation of composer Yao Ming, we use "Yao Ming_composer " to represent him, such as:
    EQL: Yao Ming_composer : place of birth :?

When the name and occupation are the same, we can continue to use more content to distinguish, such as:
    1. Song Jia_actress_after80s  default alias:junior Song Jia
    2. Song Jia_actress_after60s
    3. Song Jia_volleyball player
  If we query first Song Jia, we can use " Song Jia" or "junior Song Jia" , recommend to use "junior Song Jia" , that will benefit other users to understand the true meaning of your EQL statement, such as:
    EQL: junior songjia : graduation school:?

Case 15 use EQL statement to calculate statistical result,  use count function
    Question: Who won both the Nobel Prize and the Academy Awards? Please count
    EQL: (?x:award):instance of: Nobel Prize
        \and
      (?x:award):instance of: Academy Awards,
     ? y = count (?x)

   ANS: x = George Bernard Shaw  ehm001001,
       Bob Dylan                ehm001002
      y = 2

Case 16 calculate statistical results with EQL statements , use avg, max functions

Question: Who won both the Nobel Prize and Oscars?
Please count the average and highest value of their Nobel prizes

EQL: (? x: award): instance of : Nobel Prize
　　　\and
　　(? x: award): instance of : Academy Awards,
　　? x: award:?y ( prize :?z1 )
　　　\and
　　?y: instance of : Nobel Prize,
　　? z2 = avg (?z1) ,
　　? z3 = max (?z1)

ANS :  x = George Bernard Shaw ehm001001,
　　　　Bob Dylan　　　　ehm001002
　　　y = Nobel Prize in Literature
　　　z1 =　118165 SEK (x = George Bernard Shaw ehm001001),
　　　　　8000000 SEK (x = Bob Dylan　　　ehm001002)
　　　z2 = 4059082.5 SEK
　　　z3 = 8000000 SEK(x = Bob Dylan　ehm001002)

Statistical functions in EQL language include: count (count), avg (average), sum (total), max (take the maximum), min (take the minimum) . Where the statistical range of statistical functions,is the spo statement include the variable inside parentheses after statistical function,and its previous spo statement.

For example, the range of avg (? z1) statistics in this case
is determined by this part of the EQL statement:
　　　(? x: award): instance of : Nobel Prize
　　　　\and
　　　(? x: award): instance of : Academy Awards,
　　　? x: award:?y ( prize :?z1 )
　　　　\and
　　　?y: instance of : Nobel Prize,

EQL system first find ?x,?y,?z1 that satisfying this part of EQL statement, then use these ?z1 to calculated average value of ?z1.

These statistical functions usually follow " = " or " ( " directly, such as " ? z2 = avg (? z1) " , so they do not need to be written in the form " \count " or " \avg " .

EQL language used "," to serparate EQL statement separately, EQL system always perform EQL statements in order,from the first statement. At the same time, the result in the previous statement like ?x,?y ,can be used in the following spo statements. The order of the variables should be ? x ,? y ,? z or  x ,? y ,? z1 ,? z2 ,?z3 , ... , ?zn , and so on, the maximum number of variables is 50 .

Case 17 using alias in query
   Question: Where is the birth place of G .B Shaw ?
   EQL : G.B.Shaw:birth place:?
   ANS: Dublin    ep1900101

Note: When this EQL statement being execute, "G.B. Shaw " as an alias is automatically replaced with "George Bernard Shaw"
        "birth place" is automatically replaced with "place of birth". In the EQL language, when an alias is used in query, the result does not change.

Case 18 fuzzy query
   When a proper alias is not found, EQL can use capability from NLP natural language understanding, automatically match the content with the closest semantic (or the most similar text) ,and sent to the user.
   Question: Where was G.E.Shaw born?
      (Suppose the user has misremembered the English name of George Bernard Shaw or entered it incorrectly)
   EQL: G.E. Shaw: birthplace:?

   ANS: do not found "G.E. Shaw" , the query you actually want is " G.B. Shaw ( George Bernard Shaw): place of birth:? "  ( y/n)

After the user enters 'y' , the EQL system gives the correct query result

    EQL: G.B. Shaw : birthplace :?
    ANS: Dublin    ep1900101

Case 19    correspondence between EQL and Lambda Calculus
   Question: Find people with children born in New York
   EQL :( ?: childern): BirthPlace: New York
   ANS: Bob Dylan                    ehm001002 ,
        Melville Arthur Feynman    ehm001359
      ...

Note: Bob Dylan had a child Jakob Dylan who was born in New York, he is also a singer.

Melville Arthur Feynman is the father of famous physicist Richard Feynman. Richard Feynman was born in Queens, New York. Melville taught Richard how to think when he was a little boy. He asked Richard to imagine that if he meet the Martians, who may ask a lot of questions about the Earth. For example, why is the Earth people need to sleep at night? Melville is a really caring and talented father.

Alonzo Church is doctoral advisor of computer science pioneer Alan Turing, in 1936, he introduced powerful lambda calculus ($\lambda$ calculus), which is called the smallest universal programming language of mathematical logic. lambda calculus play a driving role in the development of computers science, and play a key role in the core of Lisp, which is an important artificial intelligence language.

Most EQL statements can be transformed into Lambda calculus expressions, that will provide strong support on mathematical and logic for EQL languages.

The EQL statement of Case 15 can be converted into a Lambda calculus statement, as follows:

EQL: (?: Childern): BirthPlace: New York
$\lambda$ calculus: $\lambda x. \exists y. Children(x, y) \wedge Birth\ Place(y, NewYork)$

After comparison, it can be seen that EQL is more concise and easier to understand than $\lambda$ calculus.

Case 20 return only part of result variables
Question: Find musicians with more than 10 children
EQL: ?x: profession : musician
\and
?x: children :?y
\and
(count (?y)> = 10),
?z = count (?y),
ANS ?x,? x.native language,?z

ANS: x = John Sebastian Bach   ehm001077 ,
x.native language = German
z = 20

.......

Note: When only part of result variables need to be returned, you can use the ANS statement to specify the returned variables. ANS is the abbreviation of answer , and you can also specify an attribute of the returned variables, such as " x.native language " . There is no ?y in the ANS statement in this case , so the result of ?y will not be returned.

Case 21 write EQL statement step by step
  Problem : find the largest U.S. states
  EQL:  ? x: instance of : U.S. states,
      ? x: area :?y,
      ? z = max(?y),
       ANS ?z
  ANS: z = 1.52 million square kilometers ( x = Alaska )

Note: to make one more comparison on EQL statements and Lambda Calculus:
  EQL:  ? x: instance of : U.S. states,
      ? x: area :?y,
      ? z = max(?y),
       ANS ?z
  $\lambda$ calculus: argmax ( $\lambda$ x. InstanceOf (x, USState), $\lambda$ x. $\lambda$ y.Area (x, y))

  In comparison, EQL sentences can be written in four steps, which is easier for ordinary people to master gradually. The lambda calculus is more suitable for genius boys and can be write down at one line.

4. Future roadmap of EQL

Case 22  Conceptual Reasoning Query (optional in first version of EQL language implementation)
  Question: Who is George Bernard Shaw's maternal grandfather?
  EQL : George Bernard Shaw : maternal grandfather : ?
  ANS: Walter Gurly   ehm001059

  Note : In EQL , some defined conceptual reasoning relationships can be used for query.

Assume that our knowledge graph only stores information about each person's parents, children , and date of birth . It does not directly store information about these people's maternal grandparents. We can do direct query by conceptual reasoning.

The definition of the "maternal grandfather" property in knowledge graph system is as follows:
   x: maternal grandfather : y
 = ( x : mother ): father : y
The system automatically converts the original query
   EQL: George Bernard Shaw :maternal grandfather : ?
Query converted to
   EQL: ( George Bernard Shaw : Mother ): Father : ?
To get the correct result :
   ANS: Walter Gurly    ehm001059

Similarly, we can define conceptual reasoning formulas for concepts such as grandfather,uncle,aunt, elder brother, yourger brother, elder sister, younger sister, etc.
For example, you can define the "elder brother " property as follows :
   x: elder brother: y
 = ( (x: father) : children : y
      \and
    (x: date of birth) > (y: date of birth)
      \and
    y: gender : male)
    \or
  ( (x: mother) : children : y
      \and
    (x: date of birth) > (y: date of birth)
      \and
    y: gender : male)

Such a complex definition can include also the half-brother of x . Such conceptual reasoning requires a large amount of calculation. Therefore, it is recommended to keep the calculated result in actual implementation, add a spo statement of " x:elder brother: y" .

Conceptual reasoning query, if fully implemented, will be more difficult and can be treat as an option for EQL version 1.0 implementation . This feature is planned to be included in EQL version 2.0 .

Case 23  natural language query by fill in the blank

One of EQL 's future development goals is to implement the natural language fill-in-the-blank operation, that is, replacing any word in natural language with  '?' ,EQL will automatically fill in the blanks and get the result.

    EQL: Walter Curly is ? of George Bernard Shaw
    ANS: maternal grandfather

    EQL: George Bernard Shaw won Nobel Prize in Literature, prize ?
    ANS: 118165 SEK

Note: We believe this fill-in-the-blank function is very suitable for using deep learning pre-training model like Google BERT . Experts in this area are welcome to make more research and development.
    For writers and white-collar workers, it is very convenient to use EQL fill-in-blanks function when writing articles , helping them quickly find the background knowledge content required for writing.

Case 24 EQL Voice Edition

    Another development roadmap for EQL is  to let user directly ask questions with their voices. EQL implements voice recognition, automatically generates EQL statement, and responds to user the query results with voice . This EQL Voice Edition can be used to enhance smart speakers.

    EQL(voice question) : What is the relationship between George Bernard Shaw and Walter Curly ?
    ANS(voice answer) : Walter Curly is maternal grandfather of George Bernard Shaw's

Case 25  suggest add request for knowledge graph

    As extreme simple query language, EQL does not recommend user to add, delete, or modify data in the knowledge graph. If the user really wishes to change the data, he can make a request for data addition, deletion , and modification through the \suggest statement . The

administrator of the knowledge graph will verify the data submitted by the user, and it will only be changed after the verification.

    Question: suggest to add data on Hemingway's main works
    EQL: \suggest add
        Ernest Hemingway : major works :  a movable feast
        (first published : 1964, ISBN: 0-684-82499-X)
    \ref1: Wikipedia   www.wikipedia.com ,
    \ref2: Ernest Hemingway Research Anthology
        ISBN 978-7-5447-3164-5
  ANS: '\suggest add' request accepted
    '\suggest add' request has been approved and the knowledge graph has been successfully updated

Note: You can use the '\suggest add' statement to request to add data to the knowledge graph. The added data is presented in the form of spo statements. In this case, the spo fact statement is :
    Ernest Hemingway : major works :  a movable feast
    (first published : 1964, ISBN: 0-684-82499-X)

  Generally, when a user submits data, the source of the data should be submitted togather for easy verification. In this case, user submitted two reference sources: Wikipedia and Ernest Hemingway Research Anthology , and submitted the address of the website and the ISBN book number, and finally passed the verification.
  Users can submit 0- n pieces of data reference source information, and the ref clause start with \ref, or \ref1, \ref2, etc.

Case 26 suggest change request for knowledge graph
  Question: suggest to change a piece of data about Ernest Hemingway's main works
  EQL: \suggest change
    Ernest Hemingway : major works : a movable feast
    (first published : 1964, ISBN: 0-684-82499-X)
  \changeTo
    Ernest Hemingway : major works : a movable feast
    (first published : 1964, ISBN: 0-684-82499-X,
     publisher: Simon & Schuster)
  \ref1 : Wikipedia www.wikipedia.com ,
  \ref2 : Ernest Hemingway Research Anthology
    ISBN 978-7-5447-3164-5

ANS: '\suggest change' request accepted

'\suggest change' request has been approved and the knowledge graph has been successfully updated

Note: You can use the '\suggest change' statement to
request data changes in the knowledge graph . You must list
the spo statement before the change , and then put the
complete spo statement after the change after \changeTo . If there is any
error in the changed spo statement , This change request will be rejected.

Case 27 suggest delete request for knowledge graph
Question: suggest delete a piece of data about Ernest Hemingway's main works

EQL: \suggest delete
Ernest Hemingway : major works : a movable feast
ANS: '\suggest delete'  request accepted
'\suggest delete' request has been rejected

Note: Generally, users should not submit a delete request, because data that is not useful to one user may be useful to other users. When submitting a delete request, you only need to submit the s:p:o part, without submitting the qv modifier clause. In this case, the customer's delete request was rejected.
If a '\suggest delete' request passes verification, at most one spo fact statement is deleted .

5. Summary
EQL, also known as Extremely Simple Query Language , can be widely used in the fields of knowledge graph, precise search, strong artificial inteligence, database, smart speaker, patent search and other fields. It can be seen from the cases of this article, that EQL adopts the principle of minimalism in design and pursues simplicity and easy to learn so that everyone can master it quickly.EQL language is obviously simpler and clearer than other data query languages such as SQL
and SPARQL , and easier to be mastered by ordinary people.

EQL language and $\lambda$ calculus are interconvertible, that reveals
the mathematical nature of EQL language,and lays a solid foundation for rigor and logical integrity of EQL language.

The underlying data storage system supporting EQL language may have multiple options, such as SQL databases, RDF databases, graph databases, MongoDB and other NoSQL databases, and even character files or Excel files. After proper processing and conversion, and development of interface software, the existing data (such as patent data, geographic information data, e-commerce data, enterprise ERP system data) can also be transformed into a data source for EQL language. In terms of application, we believe EQL can be easily applied to various industries such as banking, telecommunications, medical, Internet, manufacturing, and service industries.

We have now completed the design of the EQL language. We will launch an implementation version of EQL with the world commonsense knowledge graph system . We welcome friends from all industries to invest and sponsor. The world commensense knowledge graph system will cover most of the concepts, nouns, commonsense and formulas in most human subjects. Starting from English and Chinese, we will eventually support the languages of all countries and all nations in the world. It is expected that there will be hundreds of millions of entities and tens of billions of fact data in this knowledge graph.

In the underlying implementation, EQL will use technologies such as NLP natural language understanding and deep learning in artificial intelligence , as well as technologies in the field of big data and search engine, to ensure high-speed and accurate information search, thereby helping users to better dig out value from data assets. At the same time, we welcome people to independently design and implement the interfaces with EQL based on various underlying storage systems, especially the interfaces with SQL databases, Excel and character file. We reserve the copyright and commercial rights of EQL , and friends are welcome to contact us for the commercial authorization.

The calling of Alan Liu is strong AI.Alan Liu is the English name of Mr.Han Liu(wechat:lh7648),the first author of this paper.I will commit all my life to the research and development of strong AI, and try to integrate AI with spirituality, science, art and business.I call this integration model as "Diamond Thinking Model" , EQL is the first spiritual product based on "Diamond Thinking Model".

The EQL language and a comprehensive knowledge graph system of the world's commonsense can together form the basis of strong AI in the future, and will make up for the shortcomings caused by the lack of understanding for the world's commonsense by the current AI system. EQL language can be used not only by humans, but also as a basic language for data query and data exchange between robots.

  The research of strong AI involves mathematics, computer science, brain science, linguistics, psychology, economics, sociology and other disciplines, which is far beyond the scope of my range of skills. I look forward to your enlightenment and cooperation, to release people's potential,to help more ordinary people to achieve their ideal, to work together for the liberation of mankind.

Appendix 2

Here is some of fact data supporting the EQL query in this paper , each of the following is a spo fact statement in knowledge graph:

1. George Bernard Shaw: award : Nobel Prize in Literature
    (date: 1925 , prize: 118165 SEK )
2. George Bernard Shaw: award: Oscar for Best Adapted Screenplay
    (winning work: Flower Girl,
     date: 1939 ,
     related items: The 11th Academy Awards)
3 .Nobel Prize in Literature: instance of: Nobel Prize
4. Nobel Prize in Literature: Nature: Literary Award
5. Bob Dylan: Awards: Nobel Prize in Literature
    ( Date: 2016 ,
     Prize: SEK 8000000 ,
     Next: Ishiguro Kazuo)
6. Bob Dylan: Awards: Academy Award for Best Original Song
    (Date: 2000 ,
     Winning entry: Things Have Changed)
7. Academy Award for Best Original Song: instance of: Academy Award
8. Oscar Award for Best Adapted Screenplay: instance of: Academy Award
9. George Bernard Shaw: Chinese name:萧伯纳
10. Bob Dylan: Chinese name: 鲍勃迪伦
11. George Bernard Shaw: Place of Birth: Dublin
12. George Bernard Shaw: Alias: G.B. Shaw
13. George Bernard Shaw : Alias: Bernard Shaw
14. Place of Birth: alias: city of birth
15. Academy Awards: Alias: Oscar